%% file: main.tex
\documentclass[a4paper, amsfonts, amssymb, amsmath, prb, reprint, showkeys, nofootinbib, twoside, onecolumn]{revtex4-1}
\usepackage[english]{babel}
\usepackage[utf8]{inputenc}
\usepackage{amsthm}
\usepackage[colorinlistoftodos, color=green!40, prependcaption]{todonotes}

\newtheorem{lem}{Lemma}
\newtheorem{prop}{Proposition}

\input{preamble}
\usepackage[pdftex, pdftitle={Article}, pdfauthor={Author}]{hyperref} % For hyperlinks in the PDF
\begin{document}
\title{Two-component Pseudovectoral Chirality Function for Tetrahedra}

\author{Haina Wang}
    \email[Correspondence email address: ]{hainaw@princeton.edu}% Your name
    \affiliation{Torquato Lab, Department of Chemistry, Princeton University, Princeton, NJ 08544, USA}

\date{\today} % Leave empty to omit a date

\begin{abstract}
Chirality, the lack of inversion symmetry, is a geometrical property critical to chemistry, biology and material sciences. In the three-dimensional Euclidean space $\mathbb{R}^3$ chriality can ususally be characterized with four-point structrual information. Various functions have therefore been proposed to quantify chirality of tetrahedra, which can be extended to other 3D objects, including molecules. However, existing functions are scalars or pseudoscalars and are unable to simultaneously possess all the desirable properties of chirality functions: detectability of chirality, inversion antisymmetry and continuity. We observe that to avoid this difficulty, any chirality function for tetrahedra must be a pseudovector with at least two components. In light of this, we propose a two-component pseudovectoral chirality function for tetrahedra that satisfies all the desirable properties. We plan to use this function to map the ``chiral zeros''  of existing pseudoscalar chirality functions and to design a microstructure descriptor for the chirality of many-body systems and multi-phase media in $\mathbb{R}^3$.
\end{abstract}

\maketitle

\section{Background and significance}\label{intro}
\subsection{Chirality functions}
The importance of chirality in chemistry, biology and material science cannot be overstated. The word ``chiral'' was first coined by Lord Kelvin: ``I call any geometrical figure, or group of points, chiral\dots if its image in a plane mirror, ideally realized, cannot be brought to coincide with itself.''\cite{Thomson1904} This definition regards chirality as a discrete property: a geometrical object is either chiral or achiral. However, in many circumstances, it is convenient, or even necessary, to regard chirality as a continuous property, i.e. an object can be more chiral or less chiral.\cite{Francl2019} Circular dichroism of proteins,\cite{Kelly2005} enantioselectivity in organic synthesis due to chiral catalyst,\cite{Bandar2012} cholesteric pitch of liquid crystals\cite{Cook2000,Wilson2001} and feasibility of models for the origin of life\cite{Bailey2000} all exhibit quantitative dependence on chirality. 

It is therefore of theoretical interest and practical significance to define ``chirality functions'' for geometrical objects, including structures of molecules and bulk materials. Such functions would enable one to distinguish objects that are mirror images (i.e. enantiomers) and measure the ''degree'' of chirality. Physical properties that depend quantitatively on chirality may then be modelled by these functions. It is apparent that some desirable properties of a chirality functions include
\begin{enumerate}
    \item ``Detectability of chirality'', or strictly achiral vanishing set: The function should vanish if and only if the input object is achiral.
    \item Inversion/reflection antisymmetry: The function should take opposite values (or at least different values) for enantiomers to distinguish between absolute configurations.
    \item Continuity: An infinitesimal change in the object shape should correspond to an infinitestimal change in the function.
    \item Translational and rotational invariance: The output of the function should be independent if the input object is translated or rotated in space. (Scaling invariance may or may not be desired depending on the physical problem of interest.)
\end{enumerate}

For the planar triangle (the simplex in the two-dimensional Euclidean space $\mathbb{R}^2$), there exist pseudoscalar functions that meet all the aforementioned requirements. As a simple example, for a triangle $\bigtriangleup ABC$ where the vertices $A,B,C$ are labelled counterclockwise in that order, a chirality function $\chi$ is defined as
\begin{equation}
    \chi(\bigtriangleup ABC)=(a-b)(b-c)(c-a)
\end{equation}
where $a,b,c$ are the lengths of the sides opposite to $A,B,C$, respectively. Since a triangle is achiral if and only if it is isosceles, one easily verifies Properties 1--3 and translational and rotational invariance. If scaling invariance is desired, $\bigtriangleup ABC$ can be scaled to unit area, then $\chi$ is applied to the scaled triangle.

In three dimensions, chirality is first encountered on the four-point level. For example, the absolute configurations of complex organic molecules in 3D can often be described by the handedness of tetrahedron-shaped ``chiral centers'', where carbon is connected to four different groups. Therefore, a general chirality function is desired for the tetrahedron, the simplex in $\mathbb{R}^3$. If such a function is available, various phenomena that depend on chirality can be modelled by coarse-graining of molecules or bulk materials into one or more tetrahedra. However, defining chirality functions for tetrahedra is much more challenging than for 2D triangles: It can be shown that real-valued functions are unable to satisfy Properties 1--3 simultaneously.\cite{Fowler2005} The reason lies in the fact that the space of tetrahedron shapes is so-called chiral connected, i.e. any chiral tetrahedron can be continuously deformed into its mirror image without passing through any intermediate achiral shape. Accordingly, due to the Intermediate Value Theorem,\cite{Rudin2006} any continuous real-valued function that assigns opposite values to enantiomeric tetrahedra inevitably assigns 0 to some chiral tetrahedra. The existence of these ``chiral zeros'' violates Property 1. 

Existing chirality functions are forced to seek compromise among Properties 1--3. For example, the Osipov-Pickup-Dunmur (OPD) chirality index, inspired by optical activity of molecules, is a continuous pseudoscalar function that permits chiral zeros.\cite{Osipov1995,Millar2005} On the other hand, an algorithm proposed by Fowler and Rassat gives $R,S$ labels to tetrahedra, in a similar spirit as the Cahn--Ingold--Prelog priority rules in organic chemistry.\cite{Fowler2006} This apparently sacrifices continuity, since a tetrahedron can switch from $R$ to $S$ upon an infinitesimal deformation. A final example is a continuous function proposed by Buda and Mislow, based on the Hausdorff distance between sets of points: the minimized Hausdorff distance between a tetrahedron and its mirror image can be used as a measure of the ``degree'' of chirality. This function varies in the range $[0,1]$ and assigns equal values to enantiomers: Antisymmetry is therefore not satisfied.\cite{Buda1992} For other scalar and pseudoscalar chirality functions, see Ref. \citenum{doi:10.1002/anie.199209891} for a comprehensive review. The incapability for real-valued functions to simultaneously possess all the desired properties lead some authors to regard quantifying chirality for tetrahedra as ``attempting the impossible''.\cite{Fowler2006} 

Our proposal is motivated by a dimensional analysis due to Weinberg and Mislow.\cite{Weinberg1997} They show that the problem of chiral zeros stems from the fact that the shape space of tetrahedra (up to translation, rotation and scaling) is 5-dimensional, whereas the shape space of \textit{achiral} tetrahedra is only 3-dimensional. If a continuous pseudoscalar chirality function $\chi$ is defined on the shape space of tetrahedra, one can show that the manifold where it vanishes is 4-dimensional, i.e. there are infinitely many more zeros of $\chi$ than there are achiral tetrahedra. The problem can potentially be avoided, therefore, if the chirality function is a vector $\vec{\chi}$ with two components. The manifold where $\vec{\chi}=\mathbf{0}$ is then 3-dimensional, and could potentially be designed to be identical to the space of achiral tetrahedra. In light of this, we show that it is indeed possible to construct two-component pseudovectoral chirality functions for tetrahedra that satisfy all the desirable properties mentioned above. 

One remark must be made here on the ``degree'' of chirality implied by any chirality function. By the degree of chirality we mean a nonnegative function that takes equal values for enantiomers. Rassat and Fowler argue that there is no purely geometric meaning of ``the most chiral tetrahedron''. Indeed, it can be easily shown that if $\alpha_0:\mathcal{T}\rightarrow [0,1]$ is a function on the tetrahedron shape space $\mathcal{T}$ satisfying Properties 1,3,4, then for any continuous function $Y:[0,1]\rightarrow [0,1]$ such that $Y(x)=0$ if and only if $x=0$, the composite function $\alpha_1\coloneqq Y\circ\alpha_0$ also satisfies Properties 1,3,4.\cite{Rassat2004} By varying $Y$, the maximum point of $\alpha_1$ can be pushed to anywhere in $\mathcal{T}$ up to the achiral limit. We believe that this observation does not rend the ``degree'' of chirality a worthless concept. Instead, this concept depends on the physical problem of interest, and different chirality-dependent physical quantities might be correlated, making some chirality measures more generally applicable than others.

\subsection{Introduction to $n$-point statistics of point configurations and multi-phase media in $\mathbb{R}^3$}
This section defines $n$-point correlation functions for point configurations and multi-phase media in $\mathbb{R}^3$, followed by a discussion of their relation with chirality. These statistical functions describe the underlying microscopic structures and have been used extensively in predicting macroscopic properties, e.g. conduction coefficient, elastic moduli, and fluid permeability.\cite{Torquato2002}  

For a point configuration of $N$ particles, the \textit{generic $n$-particle probability density function} $\rho_n(\mathbf{r}^n)$ is proportional to the probability density of finding any subset of $n$ particles with configuration $\mathbf{r}^n=\{\mathbf{r}_1,...,\mathbf{r}_n\}$ in a small volume element $d\mathbf{r}^n$. It is defined as
\begin{equation}
    \rho_n(\mathbf{r}^n)\coloneqq\frac{N!}{(N-n)!}\int_{\mathbf{r}^n} P(\mathbf{r}^N) d\mathbf{r}^{N-n}
\end{equation}
where $P(\mathbf{r}^N)$ is the probability density of finding particle 1 at $\mathbf{r}_1$, and particle 2 at $\mathbf{r}_2$,..., and particle $n$ at $\mathbf{r}_n$. The volume integral of $\rho_n$ is not unity but $N!/(N-n)!$. It is convenient to define the \textit{$n$-particle correlation function}
\begin{equation}
    g_n(\mathbf{r}^n)=\rho^{-n}\rho_n(\mathbf{r}^n)
\end{equation}
where $\rho$ is the number density.

For a multi-phase media, the \textit{$n$-point probability function} $S^{(i)}_n$ of phase $i$, is defined as the probability density of finding $n$ points at positions $\mathbf{r}^n=\{\mathbf{r}_1,...,\mathbf{r}_n\}$ all in phase $i$, i.e.
\begin{equation}
    S^{(i)}_n(\mathbf{r}^n)\coloneqq\prod_{j=1}^n\mathcal{I}^{(i)}(\mathbf{r}_j)
\end{equation}
where $\mathcal{I}^{(i)}$ is an indicator function
for phase i, defined by
\begin{equation}
    \mathcal{I}^{(i)}(\mathbf{r})\coloneqq
    \begin{cases}
    1 & \mathbf{r} \in \mathcal{V}_i \\
    0 & \text{otherweise}
    \end{cases}
\end{equation}
where $\mathcal{V}_i$ is the region of phase $i$. In the following discussions, we use the term ``$n$-point correlation function'' to mean either $g_n$ and $S_n$ when the context is clear.

In the following discussions we restrict our attention to statistically homogeneous and isotropic systems, i.e. $g_n$ or $S_n$ is independent from the absolute position and orientation of the $n$-point configuration $\mathbf{r}^n$. In $\mathbb{R}^3$, chirality of such systems first occurs on the four-point level. For a four-point configuration $T=\mathbf{r}^4$ and its mirror image $T'$, if $g_4(T)\ne g_4(T')$, then the underlying many-particle system is chiral. The same can be said about $S_4$ of multi-phase media. However, four-point correlation functions are challenging to compute and have not been subject to intensive study. Partly because of this, microstructure descriptors for chirality have not been rigorously defined. In this proposal, we will show that a properly defined chirality function for tetrahedra enables the design of such a descriptor, which may find wide applications in the study of chiral photonic materials,\cite{Kragt2019,Zannotti2017}, solid catalysts in asymmetric synthesis\cite{Lin2005}, etc.

\section{Specific goals}
The three goals of this proposed study are
\begin{itemize}
    \item To explicitly construct two-component pseudovectoral chirality functions for tetrahedra that satisfy Properties 1--4 simultaneously;
    \item To map and classify the chiral zeros of previously proposed pseudoscalar chirality functions;
    \item To design a microstructure descriptor for the chirality of many-body systems and multi-phase media in $\mathbb{R}^3$;
\end{itemize}

\section{Theory, Experimental design and feasibility}
\subsection{A proposed chirality function}
 Here, we propose a chirality function $\vec{\chi}$ that satisfies Properties 1--4 above, including scale invariance. The idea behind this function is the observation that a nonplanar tetrahedron is achiral if and only if it has two isosceles faces sharing the same base or two congruent faces that are superimposable when folded about their common edge. Folding two faces about their common edge proves to be crucial to our definition of $\vec{\chi}$.
	
\begin{figure*}
		\centering
		\includegraphics[width=0.7\linewidth]{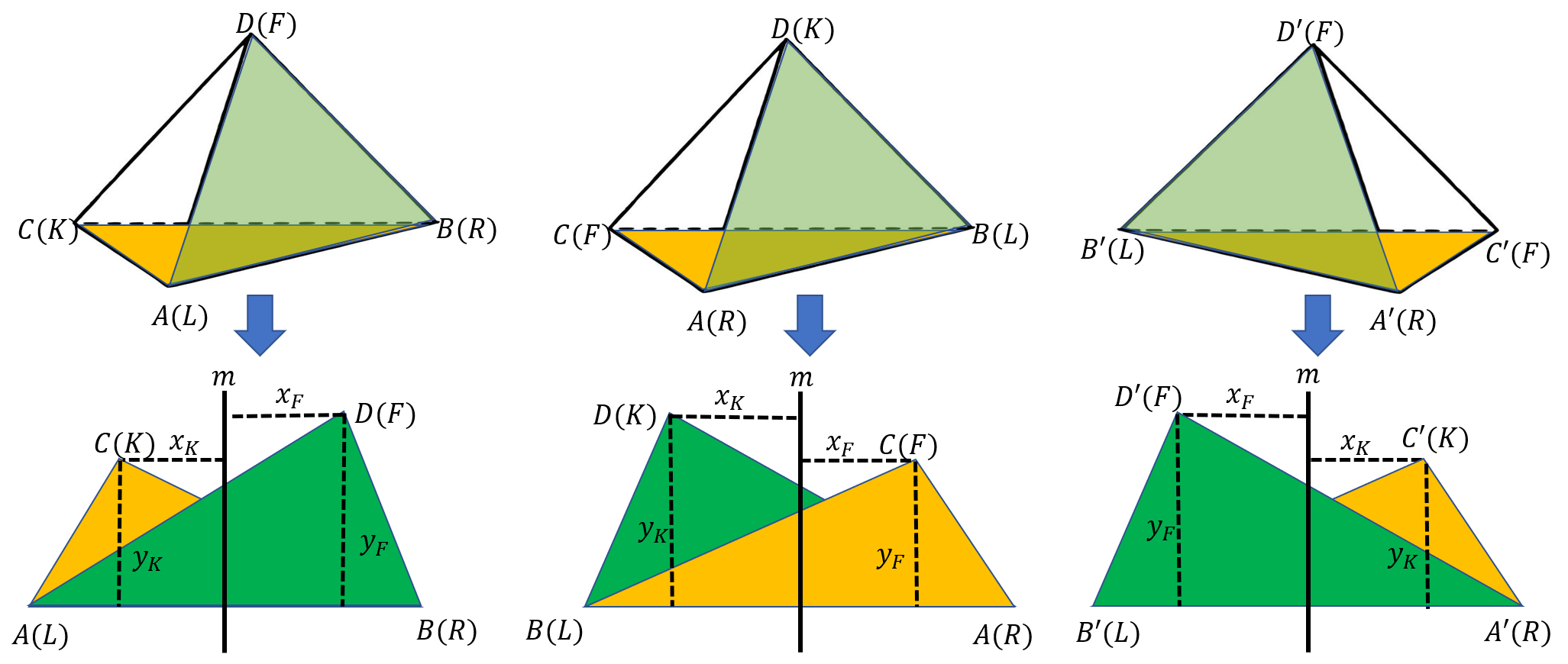}
		\caption{Treatment of the tetrahedron $ABCD$ in computing the vector $\mathbf{v}_{AB}$ (Eq. \ref{vab}). Left and middle columns: Two ways of relabelling the tetrahedron $ABCD$, and planar figures after folding the faces $ABC$ and $ABD$ about $AB$, according to the labellings in the top row. Right column: A relabelling of the mirror image $A'B'C'D'$ of $ABCD$, and planar figure after folding the faces $A'B'C'$ and $A'B'D'$ about $A'B'$.}
		\label{fig:chiralityvec}
	\end{figure*}

Let $T=ABCD$ be a tetrahedron of unit volume and let $AB$ be any edge. Relabel the vertices as $L, R, K, F$ temporarily (for left, right, back and front), such that $(\vec{LK}\times \vec{LF})\cdot\vec{LR}\geq 0$, where $L$ and $R$ are $A$ or $B$. As shown in the top row of Fig. \ref{fig:chiralityvec}, there are two ways of relabelling, but we will see shortly that both ways give the same chirality function. 
	
	Now, fold the faces $LRK$ and $LRF$ about $LR$ onto the same plane as shown in the bottom row of Fig. \ref{fig:chiralityvec}. For the planar figure $LRKF$, let $m$ be the perpendicular bisector of $LR$. Let $(x_K,y_K)$ and $(x_F,y_F)$ be the coordinates of $K$ and $F$ in the planar coordinate system established by $LR$ and $m$. That is, $x_K$ is the distance from $K$ to $m$, negative if $K$ is closer to $L$ than to $R$, and $y_K$ is the distance from $K$ to $LR$ (always nonnegative). The same is for $F$. Now, we define the vector $\mathbf{v}_{AB}$, which can be regarded as the contribution of the edge $AB$ to the chirality of $T$
	\begin{equation}
		\mathbf{v}_{AB}\coloneqq\sin(\alpha)(x_F-x_K,(y_F-y_K)(x_F+x_K))
		\label{vab}
	\end{equation}
    where $\alpha$ is the dihedral angle between the planes $ABC$ and $ABD$ in the original tetrahedron ($0\leq\alpha\leq\pi$). It can be easily seen that both labellings in Fig. \ref{fig:chiralityvec} produce the same $\mathbf{v}_{AB}$, and that $\mathbf{v}_{AB}$ is a continuous function of the tetrahedron shape. 
    
\begin{figure}
		\centering
		\includegraphics[width=0.3\linewidth]{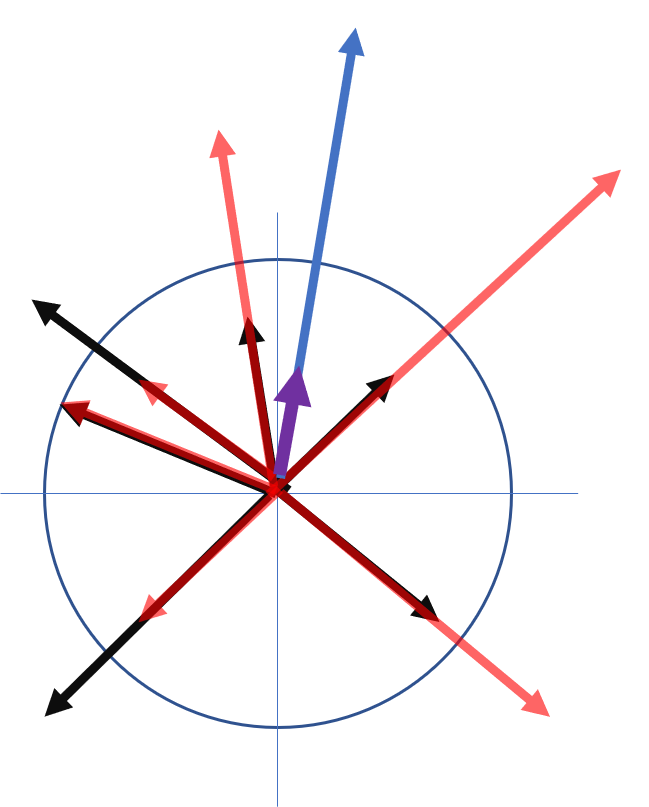}
		\caption{Illustration for computing the vector $\vec{\chi}$ in Eq. (\ref{chit}). Black: $\mathbf{v}_{PQ}$; Red: $I(\mathbf{v}_{PQ})$; Blue: $\sum_{P\ne Q}I(\mathbf{v}_{PQ})$; Purple: $I\left(\sum_{P\ne Q}I(\mathbf{v}_{PQ})\right)$. }
		\label{fig:chiralitygeominv}
	\end{figure}

    Our proposed chirality function $\vec{\chi}(T)$ is then defined as the ``harmonic mean'' of the contributions from all 6 edges (Fig. \ref{fig:chiralitygeominv}).
	\begin{equation}
		\vec{\chi}(T)=(\chi_1,\chi_2)\coloneqq I\left(\sum_{P\ne Q}I(\mathbf{v}_{PQ})\right)
		\label{chit}
	\end{equation}
	where $P,Q$ are labels of vertices and
	\begin{equation}
		I(\mathbf{w})\coloneqq \frac{\mathbf{w}}{\left|\mathbf{w}\right|^2}
	\end{equation}
	is the geometric inversion of a vector with respect to the unit circle. $I(\mathbf{0})$ is defined to be $(\infty,\infty)$, and $I((\infty,\infty))$ is defined to be $\mathbf{0}$.
	
	We now show that $\vec{\chi}$ is a function of the tetrahedron shape that satisfies Properties 1--4 in Section \ref{intro}, therefore is a proper chirality function. Continuity and translational, rotational and scaling invariance are clear from the definition. It remains to prove Properties 1 and 2 (detectability of chirality and antisymmetry). Firstly, we show that $\mathbf{v}_{AB}$ is the contribution of the edge $AB$ to the chirality of $T$ in the sense of the following lemmas.
	\begin{lem}
		If $\mathbf{v}_{AB}=\mathbf{0}$, $T$ is achiral.
	\end{lem}
	\begin{proof}
		If $\mathbf{v}_{AB}=\mathbf{0}$, one of the following statements is true: (a) $\sin\alpha = 0$; (b) $x_F=x_K$ and $y_F=y_K$; (c) $x_F=x_K=0$. Case (a) implies that the vertices are coplanar, so $T$ is trivially achiral. Case (b) implies that the faces $ABC$ and $ABD$ are superimposable if folded about $AB$, so $T$ has a mirror plane containing $AB$. Case (c) implies that $ABC$ and $ABD$ are isosceles with the same base $AB$, so $T$ has a mirror plane containing $CD$.
	\end{proof}

	\begin{lem}
		Let $T'=A'B'C'D'$ be the enantiomer of $ABCD$, where $A$ is corresponds to $A'$,etc. We have $\mathbf{v}_{A'B'}=-\mathbf{v}_{AB}$.
	\end{lem}
	\begin{proof}
		In Fig. \ref{fig:chiralityvec}, the middle and right columns show that reflection of $ABCD$ amounts to interchanging the labels $K$ and $F$. The lemma follows from Eq. (\ref{vab}).
	\end{proof}

	\begin{prop}
		Let $T$ be a tetrahedron of unit volume. $\vec{\chi}(T)$ vanishes if and only if $T$ is achiral. Furthermore, for enantiomers $T$ and $T'$, we have $\vec{\chi}(T)=-\vec{\chi}(T')$.
	\end{prop}
	\begin{proof}
		If $T$ is achiral, at least one of the $\mathbf{v}_{PQ}$ vanishes, so $\vec{\chi}(T)=\vec{0}$. Conversely, if $\vec{\chi}(T)=\mathbf{0}$, $\sum_{P\ne Q}I(\mathbf{v}_{PQ})=(\infty,\infty)$, so at least one of the $I(\mathbf{v}_{PQ})$ must be infinite, and the corresponding $\mathbf{v}_{PQ}$ vanishes, implying an achiral tetrahedron. Reflection of $T$ into its enantiomer negates all the $\mathbf{v}_{PQ}$, and geometric inversion preserves the vector direction. Therefore, $\vec{\chi}(T)=-\vec{\chi}(T')$.
	\end{proof}
	
	\subsubsection*{Feasibility}
\begin{figure}
		\centering
		\includegraphics[width=0.5\linewidth]{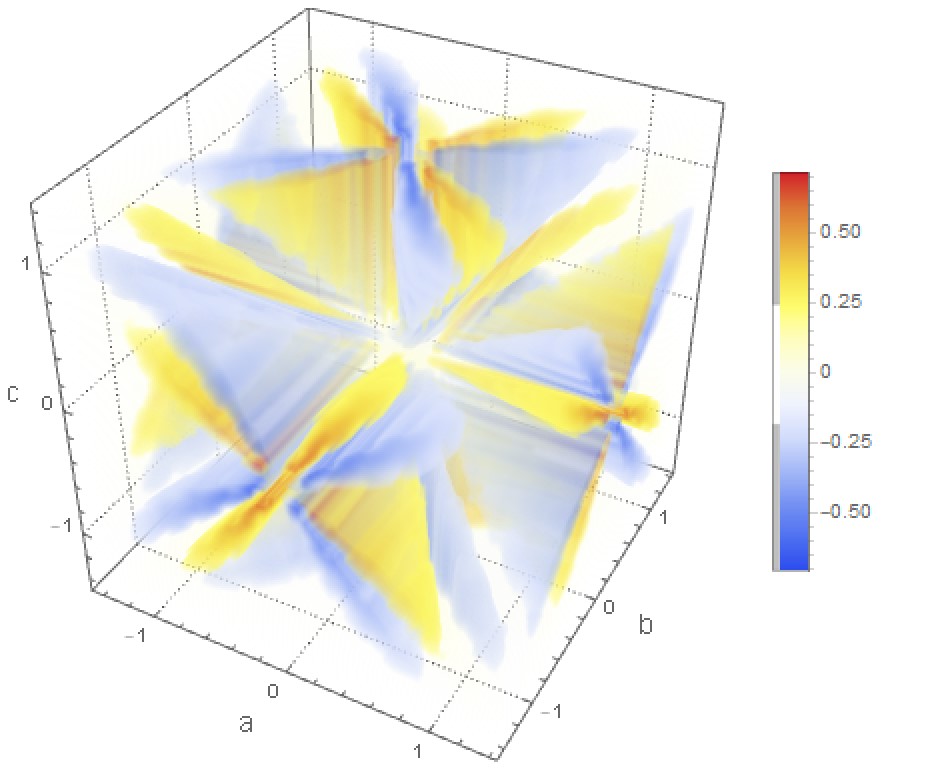}
		\caption{Plot of $\chi_2$ (Eq. (\ref{chit})) of rectangular tetrahedra. The vertices are at $A(a,0,0), B(0,b,0), C(0,0,c), D(0,0,0)$.}
		\label{fig:rectangularty}
	\end{figure}

\begin{figure}
		\centering
		\includegraphics[width=0.5\linewidth]{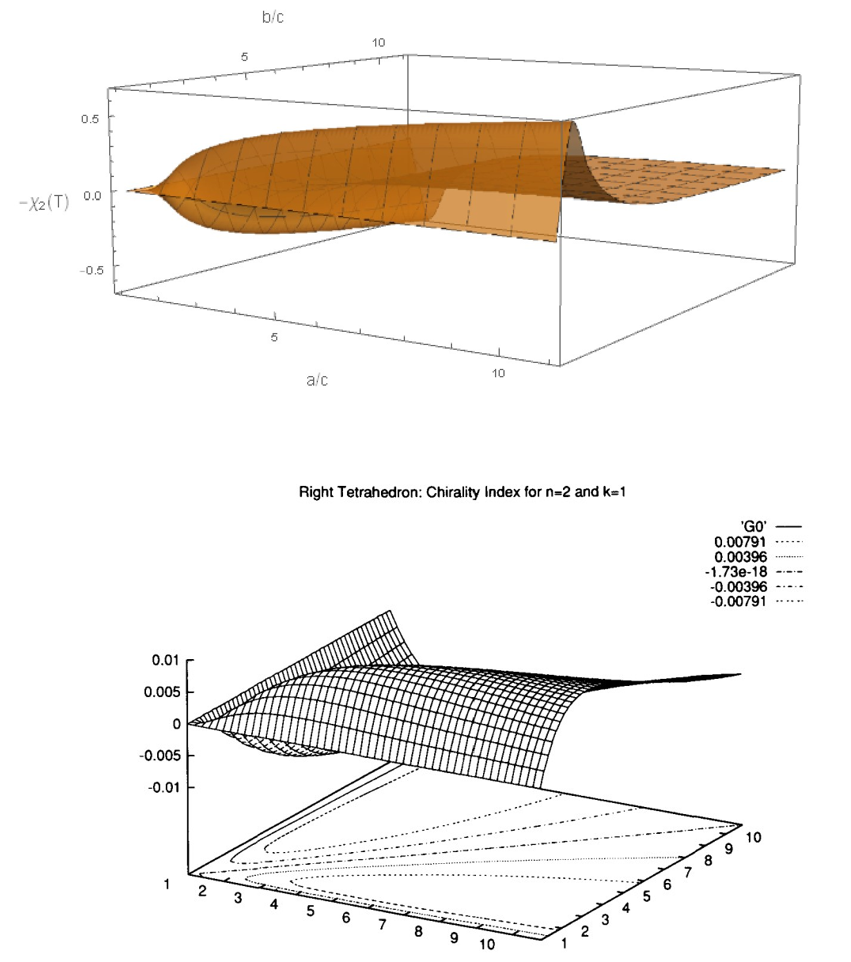}
		\caption{Top: Plot of $-\chi_2$ of rectangular tetrahedra against $a/c$ and $b/c$. Bottom: (Reproduced from Ref. 
		\citenum{Osipov1995}) Plot of the OPD index for rectangular tetrahedra against $a/c$ and $b/c$.}
		\label{fig:rectangularCompare}
	\end{figure}

\begin{figure}
		\centering
		\includegraphics[width=0.5\linewidth]{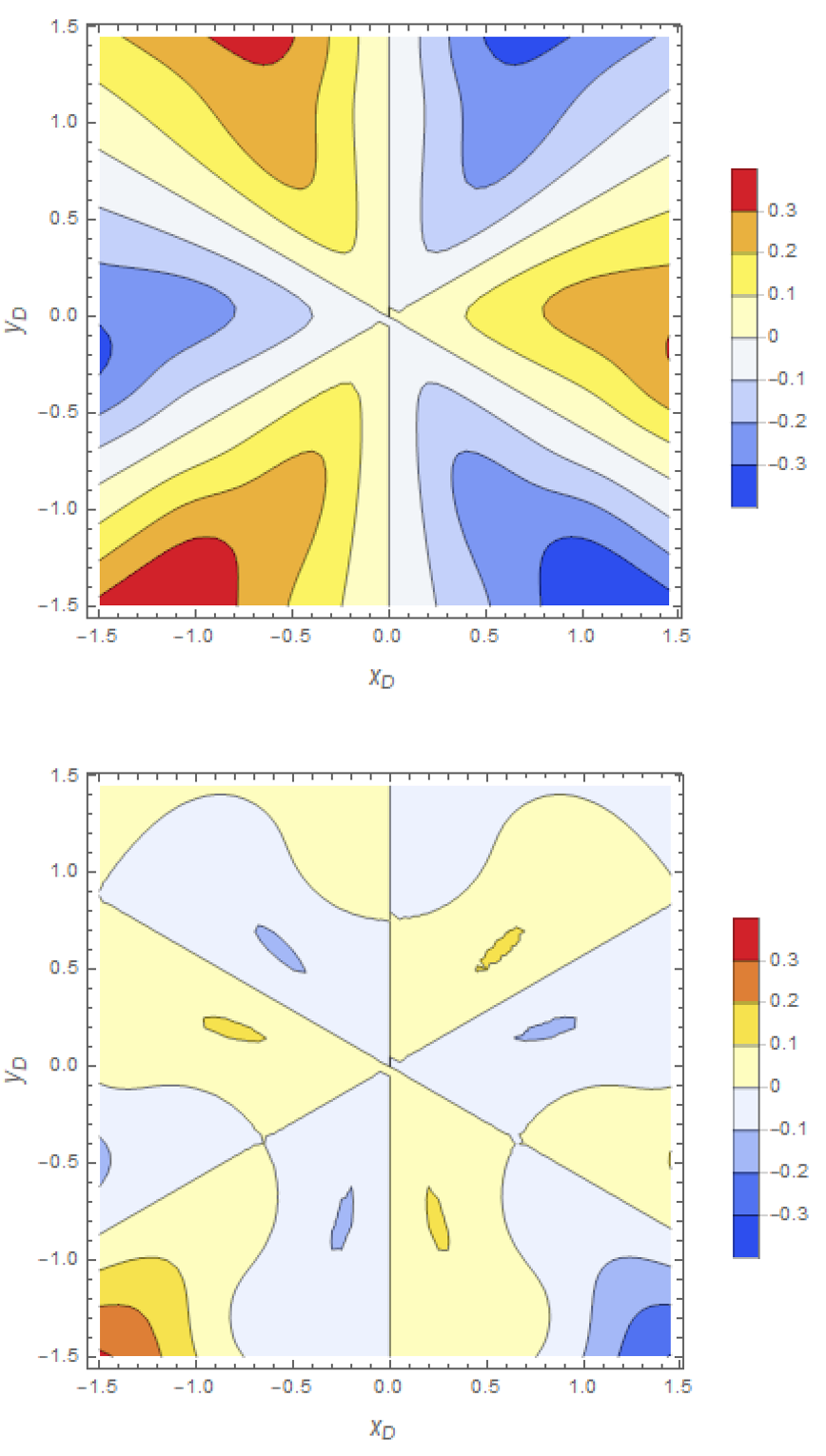}
		\caption{Plot of $\vec{\chi}$ of some tetrahedra with a regular triangular base. The vertices are at $A(-\sqrt{3}/2,1/2,0),B(\sqrt{3},1/2,0),C(0,1,0),D(x_D,y_D,\sqrt{6}/3)$. Top: $\chi_1$; Bottom: $\chi_2$. The axes are $x_D,y_D$.}
		\label{fig:rightTri}
	\end{figure}

	I have programmed Eq. (\ref{chit}) and computed $\vec{\chi}(T)$ for rectangular tetrahedra, for which the vertices are positioned at $A(a,0,0), B(0,b,0), C(0,0,c), D(0,0,0)$. Because the opposite edges for these tetrahedra are always perpendicular, $\chi_1$, the first component of $\vec{\chi}$, vanishes. Fig. \ref{fig:rectangularty} plots $\chi_2$ against $a,b,c$, where one could immediately notice the inversion and reflection antisymmetry. Fig. \ref{fig:rectangularCompare} compares $-\chi_2$ and the OPD index for rectangular tetrahedra in Ref. \citenum{Osipov1995}. We observe that they are qualitatively similar. $\vec{\chi}$ decays to zero faster due to the nature of the harmonic mean in Eq. \ref{chit}. Since the OPD index is related to optical activity, the similarity of $\vec{\chi}_2$ with OPD implies that $\vec{\chi}$ could be a useful chirality function in practice. Fig. \ref{fig:rightTri} plots $\vec{\chi}$ for a class of tetrahedra with a regular triangular base. The vertices are at $A(-\sqrt{3}/2,1/2,0)$, $B(\sqrt{3},1/2,0)$, $C(0,1,0)$, $D(x_D,y_D,\sqrt{6}/3)$. Again, it is clear that for the tetrahedrons represented in this figure, $\vec{\chi}(T)=0$ if and only if the tetrahedron is achiral.Computational cost for $\vec{\chi}$ is tiny: each $\vec{\chi}$ evaluation took no more than $3\times 10^{-6}$ CPU second on a 2.5 GHz processor. 
	
	One potential pitfall for our definition of $\vec{\chi}$ is that could diverge if $\sum_{P\ne Q}I(\mathbf{v}_{PQ})=\mathbf{0}$. It is not known at the moment whether this really happens for some tetrahedra. This question could be answered by mathematically establishing the relationship among $I(\mathbf{v}_{PQ})$, or by numerically maximizing $\vec{\chi}$ over the tetrahedron shape.
	
	\subsection{Mapping chiral zeros of existing pseudoscalar chirality functions}
	Our newly defined pseudovectoral function $\vec{\chi}$ is able to map existing pseudoscalar chirality functions. In particular, chiral tetrahedron shapes that make existing functions vanish can be clearly identified as chiral by $\vec{\chi}$, and interesting information may be revealed. We use the OPD chirality index (denoted $G_0(T)$) as an example to describe our procedure. It is defined by
	\begin{equation}
	    G_0(T)=\sum_{ijkl}\left(\mathbf{u}_{ij}\cross \mathbf{u}_{kl}\cdot \mathbf{u}_{il}\right)\left(\mathbf{u}_{ij}\cdot\mathbf{u}_{jk}\right)\left(\mathbf{u}_{jk}\cdot\mathbf{u}_{kl}\right)
	\end{equation}
	where $i,j,k,l$ are vertices $A,B,C,D$, $\mathbf{u}_{ij}\coloneqq\vec{ij}/\left|ij\right|$ when $i\ne j$, and $\mathbf{u}_{ij}\coloneqq 0$ when $i=j$. We aim to identify the chiral zeros of $G_0$ and map their $\vec{\chi}$ values on $\mathbb{R}^2$.
	
\begin{figure}
		\centering
		\includegraphics[width=0.5\linewidth]{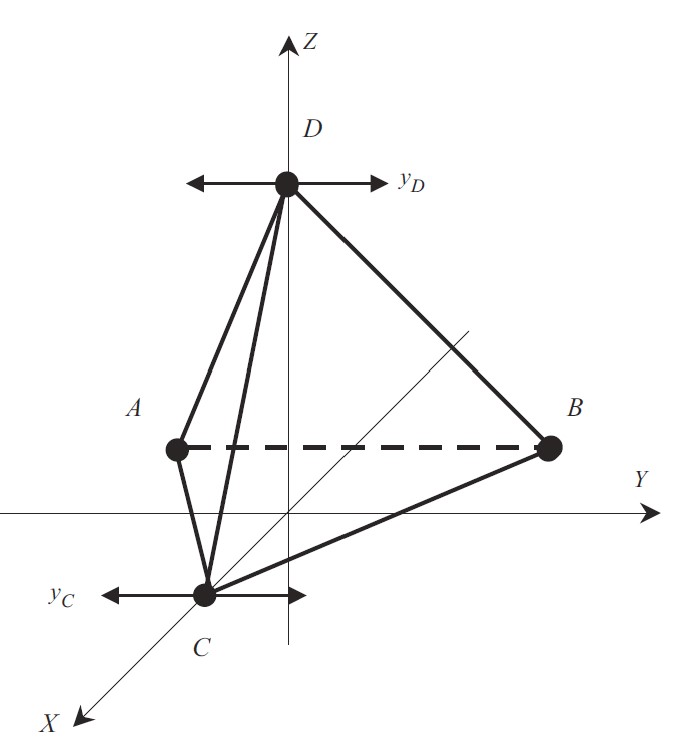}
		\caption{(Reproduced from Ref. \citenum{Millar2005}) Chiral transformations of scaled tetrahedra: The length of edge $AB$ is set equal to 1. Vertices $A$ and $B$ are fixed at points $(-0.25, -0.5, 0)$ and $(-0.25, 0.5, 0)$, respectively. Generally, vertex $C$ can move freely in the $XY$ plane and vertex $D$ can move in three dimensions. To reduce the dimensionality of the problem, when generating the $G_0$ surface, vertices $C$ and $D$ are allowed to move only in the $y$ direction, so that their coordinates are $(0.25, y_C, 0)$ and $(0, y_D, 0.25)$, respectively.}
		\label{fig:dimensions}
	\end{figure}

	The shape space of tetrahedra is 5-dimensional and will be characterized in the manner following Ref. \citenum{Millar2005} (Fig. \ref{fig:dimensions}). Let $A,B$ be two vertices of tetrahedron $T$. By appropriate scaling, translation and rotation, we fix $A$ and $B$ at $(-0.25,-0.5,0)$ and $(-0.25,0.5,0)$, respectively, and fix $C$ in the $xy$-plane. The tetrahedron shape $T$ is then described by five coordinates $x_C,y_C,x_D,y_D,z_D$. To further reduce dimensionality, we fix $x_C,x_D,z_D$ and plot $G_0(T)$ against $x_D,y_D$ (See Fig. \ref{fig:OPD}. The intersection curve of the $G_0$ surface with the surface $G_0=0$ is a subset of zeros of $G_0$. Because the space of achiral tetrahedra is 3-dimensional, on this curve at most a finite number of points correspond to an achiral tetrahedron, and the other points correspond to chiral zeros. The pseudovectoral chirality function $\vec{\chi}(T)$ is then computed for the chiral zeros and plotted onto a 2D plane. The $\vec{\chi}$-image of these chiral zeros will be a centralsymmetric curve on $\mathbb{R}^2$ about the origin. Other combinations of $x_C,x_D,z_D$ are chosen and the corresponding chiral zeros are identified and mapped.
	
	The $\vec{\chi}$-image of all chiral zeros for $G_0$  will be either a set of centralsymmetric curves, at least one of which passing through the origin (i.e. true achirality), or a centralsymmetric region containing the origin. This map can potentially reveal interesting insights into the OPD chirality index. For example, if the image of chiral zeros occupy only limited regions in $\mathbb{R}^2$, one could determine the class of tetrahedrons that are free from chiral zeros, i.e. for which the OPD chirality index is practically applicable. The chiral zeros of other pseudoscalar chirality functions can be studied the same way. 
	
	We remark that if for two pseudoscalar chirality functions $G$ and $H$, their $\vec{\chi}$-images on $\mathbb{R}^2$ are simple curves intersecting only at the origin, then $G$ and $H$ can be used to define another two-component pseudovectoral chirality function. 
	
    \subsubsection*{Feasibility}
\begin{figure}
		\centering
		\includegraphics[width=0.5\linewidth]{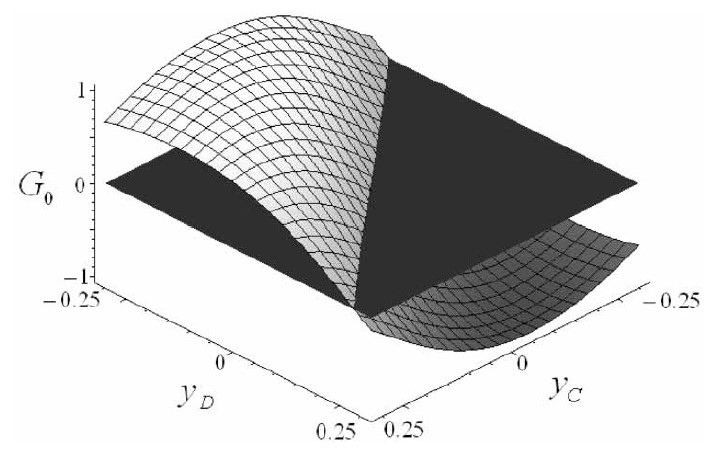}
		\caption{(Reproduced from Ref. \citenum{Millar2005}) $G_0$ surface for scaled tetrahedra of Fig \ref{fig:dimensions}. The dark horizontal plane corresponds to $G_0=0$.}
		\label{fig:OPD}
	\end{figure}

    The aforementioned plot of $G_0$ against $x_D,y_D$ has been plotted for a given set of $(x_C,x_D,z_D)=(0.25,0,0.25)$, and a set of chiral zeros has been identified (Fig. \ref{fig:OPD}). Therefore, our goal described above is essentially repetition of this procedure for other sets of $(x_C, x_D, z_D)$ and is readily achievable. The computational challenge is equivalent to evaluating $G_0$ and $\vec{\chi}$ for about $20^5$ tetrahedron shapes (assuming a grid of 20 points in each dimension), which can be accomplished in minutes on a 2.5 GHz processor. 
    
    One potential pitfall is that the $\vec{\chi}$-image of chiral zeros of $G_0$ may turn out to cover all of $\mathbb{R}^2$. As a result, no insightful information can be deduced about the relation between these chirality functions. If this happens to be the case, attention must be restricted to subspaces of tetrahedron shapes with certain symmetries, e.g. $C_2$ or $D_2$. 
    
    \subsection{Chirality distribution functions for many-body systems and multi-phase media}
    
    As mentioned in Introduction, in many scenarios it is important to describe chiral properties of many-body systems and multi-phase media in $\mathbb{R}^3$. Our chirality function $\vec{\chi}$ could potentially provide a powerful microstructure descriptor that captures chirality information for such systems. This subsection defines this microstructure descriptor, which we term \textit{four-point chirality distribution function} and denote by $f_\chi(\mathbf{y})$, where $\mathbf{y}\in\mathbb{R}^2$. 
    
    For a single point configuration in $\mathbb{R}^3$ with $N$ particles, $f_\chi(\mathbf{y})$ is the probability density of finding four particles in a tetrahedron configuration whose chirality function $\vec{\chi}$ lies in a small neighborhood of $\mathbf{y}$. In other words, $f_\chi(\mathbf{y})$ is related to the four-particle correlation function $g_4$ by
    \begin{equation}
        f_\chi(\mathbf{y})\coloneqq\frac{(N-4)!}{N!}\rho^4\int g_4(\mathbf{r}^4)\delta\left[\vec{\chi}(\mathbf{y}-\mathbf{r}^4)\right]d\mathbf{r}^4
        \label{fchi}
    \end{equation}
    where $\mathbf{r}^4=\{\mathbf{r}_1,\mathbf{r}_2,\mathbf{r}_3,\mathbf{r}_4\}$ refers to positions of four particles. In many situations, we are interested only in the chirality of four-particle configurations with a certain property $P$, e.g. the tetrahedron they form is smaller than a certain volume, or such that the longest edge is smaller than a certain length. In such cases, we define the chirality distribution function subject to the property $P$ as
    \begin{equation}
        f_\chi(\mathbf{y};P)\coloneqq\frac{\int_{\mathbf{r}^4\in P} g_4(\mathbf{r}^4)\delta\left[\mathbf{y}-\vec{\chi}(\mathbf{r}^4)\right]d\mathbf{r}^4}{\int_{\mathbf{r}^4\in P} g_4(\mathbf{r}^4) d\mathbf{r}^4}\\
        \label{fchiP}
    \end{equation}
    In other words, $f_\chi(\mathbf{y};P)$ is defined as the conditional probability density of finding a four-particle configuration whose chirality function $\vec{\chi}$ is in a small neighborhood of $\mathbf{y}$, given that the configuration satisfies $P$. We assume here that $P$ is a set of nonzero measure.
    
    For a multi-phase media, the four-point chirality distribution function $f_\chi^{(i)}(\mathbf{y})$ of phase $i$ is defined as the conditional probability density of finding four randomly placed points in a configuration whose chirality function $\vec{\chi}$ is in a small neighborhood of $\mathbf{y}$, given that the four points are all in phase $i$. It is related to the four-point probability function $S_4$ by
    \begin{equation}
        f_\chi^{(i)}(\mathbf{y})\coloneqq\frac{\int S_4(\mathbf{r}^4)\delta\left[\mathbf{y}-\vec{\chi}(\mathbf{r}^4)\right]d\mathbf{r}^4}{\int S_4(\mathbf{r}^4) d\mathbf{r}^4}\\
        \label{fchii}
    \end{equation}
    if only four-point configurations with a certain property is of interest, the integrals are made over those configurations, as in Eq. (\ref{fchiP}).
    
    We will focus on the chirality distribution function for point configurations (Eq. (\ref{fchi}) and (\ref{fchiP})) in following discussions, while keeping in mind that the treatment for multi-phase media is exactly parallel. If the point configuration is \textit{statistically achiral on the four-point level}, i.e. any four-point configuration is as likely as its mirror image in the infinite volume limit, then $f_\chi(\mathbf{y};P)=f_\chi(\mathbf{-y};P)$ for any property $P$ that does not involve chirality. On the other hand, for a \textit{statistically chiral} configuration on the four-point level, there exists some $P$ not related to chirality such that $f_\chi(\mathbf{y};P)$ is not centrally symmetric. 
    
    \subsubsection*{Algorithm for computing $f_\chi$}
    
    Extracting $f_\chi$ from experimental or simulated data is straightforward. We describe here the procedure of computing $f_\chi(\mathbf{y};P)$ for a single point configuration under periodic boundary conditions.
    
    \begin{enumerate}
        \item Pixelate $\mathbb{R}^2$ with the origin at the center. From current experience with other distribution functions, pixel size of at most $0.1\times0.1$ with at least 50$\times$50 pixels is a minimal requirement to obtain interesting information. It may also be useful to pixelate $\mathbb{R}^2$ with polar coordinate: a typical pixel would be 0.1$\times 7.2^{\circ}$. Establish a counter for every pixel, initialized to 0.
        \item Under periodic boundary conditions and minimal image approximation, identify all four-point configurations that satisfy $P$. Let $M$ be the number of such configurations. 
        \item Compute $\vec{\chi}(\mathbf{r}^4)$ for every four-point configuration above and add 1 to the counter of the pixel corresponding to $\vec{\chi}(\mathbf{r}^4)$.
        \item Divide the counter of every pixel by  $MA_p$, where $A_p$ is the area of the pixel. This gives $f_\chi(\mathbf{y};P)$.
    \end{enumerate}
    To compute $f_\chi^{(i)}(\mathbf{y};P)$ of a multi-phase media, all steps remain the same, except that Step 2 involves randomly generating four-point configurations such that all points are in phase $i$ and $P$ is satisfied. To do this, a Poisson point pattern is superimposed with the 3D material image. The points that lie in phase $i$ are subject to the same analysis as Step 2 above.
    
    \subsubsection*{Computer experiments to test the applicability of $f_\chi$}
    Here, we describe computer experiments to test whether $f_\chi$ defined above is a robust and useful descriptor for chirality of 3D materials in general.
    
    First, we test the behavior of $f\chi(\mathbf{y},P)$ for known achiral states of matter. Benchmark point configurations and two-phase media are prepared in $\mathbb{R}^3$ under periodic boundary conditions, which include
    \begin{itemize}
        \item A $20\times20\times20$ integer lattice with number density $\rho=0.5$.
        \item An equilibrium Lennard-Jones supercritical fluid with $N=8000$. The configuration is obtained from Monte-Carlo simulation under the pair potential $v(r)=4(r^{-12}-r^{-6})$ with cutoff $r=2.5$, $\rho=0.5$ and temperature $T=2.0T_c$, where $T_c$ is the critcal temperature.
        \item A Poisson point pattern at $\rho=0.5$.
        \item A two-phase media obtained by decorating a $10\times10\times10$ integer lattice with spheres of unit diameter centered at the lattice points, such that the packing fraction is $\phi_i=0.48$.
        \item An equilibrium packing of 1000 hard spheres of unit diameter at $\phi_i=0.48$, obtained by Monte-Carlo simulation. (This is below the freezing packing fraction.)
        \item A $10\times10\times10$ random checkerboard of unit grid volume at $\phi_i=0.48$.
    \end{itemize}
    Let $P_l$ be the condition that the longest edge of the four-point configuration is smaller than $l$. The chirality correlation function $f_\chi(\mathbf{y};P_l)$ or $f_\chi^{(i)}(\mathbf{y};P_l)$ is computed for each of these systems with the procedure above and for $l=2,3,4$. For the two-phase media, $f_\chi^{(i)}(\mathbf{y};P_l)$ is computed by sampling the system with a Poisson point configuration with $N=16000$, i.e. about 8000 points lie in each phase. Due to the definition of $P_l$, it is not necessary to enumerate all four-point configurations. Instead, a \textit{neighbor-list} is established that contains the positions of all particles in a sphere of radius $l$ centered at each particle. Only four-point configurations formed by particles in the spheres are considered, because only those could possibly satisfy $P_l$. We expect that the chirality density functions of all systems above are centrally symmetric, i.e. $f_\chi(\mathbf{y};P_l)-f_\chi(-\mathbf{y};P_l)$ is expected to be on the order of random errors for all $\mathbf{y}$.
    
    Next, we test the behavior of $f_\chi(\mathbf{y},P)$ for known chiral states of matter. The benchmark systems are
    \begin{itemize}
        \item A perturbed $20\times20\times20$ integer lattice with number density $\rho=0.5$, where the perturbation is spiral: lattice point $(x,y,z)$ is perturbed to $(x+0.1\cos z, y+0.1\sin z,z)$.
        \item Same as above, but with the reverse spiral perturbation: $(x,y,z)\rightarrow(x-0.1\cos z, y-0.1\sin z,z)$.
        \item Two-phase media obtained by decorating the above two systems with spheres of unit diameter centered at the lattice points, such that the packing fraction is $\phi_i=0.48$. A 1000-particle subset is then used for more effective computation.
    \end{itemize}
    The chirality correlation function $f_\chi(\mathbf{y};P_l)$ (or $f_\chi^{(i)}(\mathbf{y};P_l)$) is computed for each of these systems and for $l=2,3,4$. They should be not centrally symmetric for at least one value of $l$.
    
    \subsubsection*{Feasibility}
    For a point configuration with $N=8000$, the number of four-point configurations is on the order of $10^{14}$. It takes on the order of years to compute $\vec{\chi}(T)$ for all the configurations, and accordingly computing $f\chi(\mathbf{y})$ is not a viable task. However, if the neighbor-list approach described above is applied to compute $f\chi(\mathbf{y};P_l)$, the computational cost can be significantly reduced. One can show that the number of four-particle configurations $N_T$ that must be considered is approximately
    \begin{equation}
        N_T=N{\frac{4\rho\pi l^3}{3}\choose 3}
    \end{equation}
    For $N=8000,\rho=0.5,l=4$, we have $N_T=3.1\times 10^{9}$. Assuming each evaluation of $\vec{\chi}$ takes $3\times10^{-6}$ s, the computational time for $f_\chi(\mathbf{y};P_8)$ for the experiments above is approximately $1.0\times10^4$ s. For $N=8000,\rho=0.5,l=2$, $N_T=5.4\times 10^{6}$. The same configuration is counted four times in $N_T$, and not all of them satisfy the condition $P_l$. However, due to the large value of $N_T$, we expect those configurations that do satisfy $P_l$ are sufficiently numerous to enable  accurate computation of the chirality distribution functions, and their
    central symmetry, or the lack of it, should be clearly observed. 

\section{Concluding remarks}
    We defined a two-component chirality function for tetrahedra that satisfy detectability of chirality, antisymmetry, continuity and translational and rotational invariance simultaneously, which can't be achieved with existing scalar or pseudoscalar chirality functions. Using this function, we also designed microstructure descriptors for the chirality of point configurations and multi-phase media at the four-point level. These functions, or possibly their modifications, could be potentially useful in modelling phenomena where macroscopic physical properties depend quantitatively on chirality of the underlying microscopic structures, provided that the shape of a dopant molecule can be reasonably coarse-grained into one or many tetrahedra.
    
    We note here that our functions $f_\chi$ describe chirality up to the four-point level. It may arrive that systems with identical $g_4$ or $S_4$ are still chiral due to difference in high-order correlation functions.\cite{Stillinger2019} Using the same dimensional analysis in Ref. \citenum{Weinberg1997}, one could show that a chirality function for $n$-point configurations in $\mathbb{R}^3$ satisfying properties 1--4 would require at least $n-2$ components. However, most chemically interesting chiral structures can be well characterized on the four-point level, and it is not advisable to define a separate chirality function on every $n$-point level. Instead, coarse-graining techniques should be used in combination of $f_\chi$ to extract as much chirality information as possible.
    
\bibliography{chirality}

\end{document}

%% file: preamble.tex
\usepackage{amsthm}
\usepackage{mathtools}
\usepackage{physics}
\usepackage{xcolor}
\usepackage{graphicx}
\usepackage[left=23mm,right=13mm,top=35mm,columnsep=15pt]{geometry} 
\usepackage{adjustbox}
\usepackage{placeins}
\usepackage[T1]{fontenc}
\usepackage{lipsum}
\usepackage{csquotes}